\documentclass[12pt]{iopart}

\usepackage{setstack}
\usepackage{iopams}
\newcommand{\eqref}[1]{(\ref{#1})}

\usepackage{xspace}
\usepackage{graphicx}
\usepackage{fullpage}

\usepackage{geometry}
\usepackage[square,sort&compress,numbers]{natbib}
\usepackage[group-separator={,}]{siunitx}
\usepackage[T1]{fontenc}

\renewcommand{\vec}[1]{\boldsymbol{#1}}

\begin{document}

\title{Universal statistics of vortex tangle in three-dimensional wave chaos}
\author{Alexander J Taylor}
\address{H H Wills Physics Laboratory, University of Bristol, 
    Tyndall Avenue, Bristol, BS8 1TL, UK}
\eads{\mailto{alexander.taylor@bristol.ac.uk}}

\begin{abstract}
  The tangled nodal lines (wave vortices) in random, three-dimensional wavefields are studied as an exemplar of a fractal loop soup. Their statistics are a three-dimensional counterpart to the characteristic random behaviour of nodal domains in quantum chaos, but in three-dimensions the filaments can wind around one another to give distinctly different large scale behaviours. By tracing numerically the structure of the vortices, their conformations are shown to follow recent analytical predictions for random vortex tangles with periodic boundaries, where the local disorder of the model `averages out' to produce large scale power law scaling relations whose universality classes do not depend on the local physics. These results explain previous numerical measurements in terms of an explicit effect of the periodic boundaries, where the statistics of the vortices are strongly affected by the large scale connectedness of the system even at arbitrarily high energies. The statistics are investigated primarily for static (monochromatic) wavefields, but the analytical results are further shown to directly describe the reconnection statistics of vortices evolving in certain dynamic systems, or occurring during random perturbations of the static configuration.
\end{abstract}

\section{Introduction}

Many physical systems exhibit tangles of filamentary loops that wind
around one another in a disordered fashion and resemble random walks on
large scales.
Examples are as diverse as $U(1)$ models of cosmic strings on a random
phase lattice~\cite{vachaspati84,hindmarsh95,strobl97}, the optical
vortices of randomly scattered coherent light~\cite{oholleran08},
closed `worm' loops of alternating spin flavours in spin
ices~\cite{jaubert11} or the Potts model~\cite{khemani12}, the
molecular filaments of a polymer melt~\cite{scherrer86,degennes79},
models of random spatial permutations~\cite{grosskinsky12}, and the
phase vortices in wave chaos that we investigate
here~\cite{taylor14,taylor17}.
In many of these cases, the bulk \emph{loop soup}
of intertwining filaments appears to take on certain universal
behaviours in large scale statistics such as the distribution of loop
lengths.
These results depend only on the isotropic random nature of the field
and not on the specific local physics.
The different behavioural regimes of ideal loop soups have recently
been fully described by work on the universal statistics of vortex
line tangles~\cite{nahum12,nahum13}, where the filaments display
distinctly different types of behaviour at different lengthscales,
depending on only a small number of parameters. Short filaments are
dominated by the local physics of the system, while the tangle on
large scales approximates an ideal scale invariant loop system. The model
also accounts for the effect of periodic boundaries, under which the
global statistics of the longest filaments are modified and the loop
soup no longer appears scale invariant.

Here we study the tangle of \emph{phase vortices} (nodal lines, phase
singularities) in three-dimensional (3D) \emph{wave chaos} arising from
linear superpositions of Gaussian random complex waves. These form a
continuous exemplar of a fractal loop soup in which there is only one
physical lengthscale, the wavelength~\cite{taylor14,taylor16}, and in
which the random statistics of the field arise only from its initial
conditions and not from extra physical processes.
The vortices can be traced across all lengthscales from the smallest
sub-wavelength structures whose geometry is dominated by the
smoothness of the wavefield, to the largest scales of a given system~\cite{taylor14,taylor16}.
Despite the linear nature of the wave interference, the vortex lines
wind around one another to form complicated random structures in which
each individual filament has been shown to take the scaling relations
of a random walk~\cite{berry00,taylor14,oholleran08}, and whose combined
bulk is thought to take an ideal scale invariant fractal distribution in
the limit of large lengthscales.
These vortex tangles are of particular interest as a 3D
counterpart to the characteristic random statistics of chaotic
eigenfunctions in ergodic cavities, which in two dimensions are well
described by random wave interference~\cite{berry77}, exhibiting
particular power law scaling relations in statistics such as nodal
domain areas and fractalities. In two dimensions this can be
understood in terms of a connection to 2D
percolation~\cite{bogomolny02,blum02} and conformal field
theory~\cite{keating06}. The tangle of vortices in three-dimensions
exhibits scaling laws resembling those in two
dimensions~\cite{vachaspati84,oholleran08,taylor14}, but with
distinctly different exponents, and it has been unclear what governs
these statistics.
We show here that all of these statistics, including certain boundary
effects that have appeared in numerical
investigations~\cite{oholleran08,taylor14}, match the predictions for
an ideal loop soup~\cite{nahum12}, a 3D counterpart to
the way many two-dimensional (2D) statistics are fixed by conformal
invariance.
Following~\cite{nahum13}, the statistics of long lines are understood
in terms of a topological effect coming from the periodic boundaries,
which change the statistics how vortices approach one another even at
arbitrarily high energies.
We further relate the statistical model for loop soup statistics
directly to the statistics of reconnections between dynamically moving
vortices, emphasising how the random distribution of reconnection
events leads to the same universal behaviours even in random
wavefields with very different parameters.

The universal scaling statistics of nodal lines in wave chaos are well
understood in the case of 2D chaotic eigenfunctions, in
which the systems are time-reversal symmetric and so can be taken as
real valued. The nodal lines are here of interest as a core structure
whose varied conformations capture the chaotic character of the
wavefunction~\cite{blum02}. Although they are confined to the plane
and so cannot truly be tangled around one another, the lines form
complicated extended conformations across all lengthscales, bounding
extended nodal domains of all-positive or all-negative value. The
shapes of the random structures are heavily influenced by their
two-dimensionality and the fact that the nodal lines generically
do not intersect~\cite{monastra03}, so the nodal lines do not behave
as Brownian random walks but instead fall into a different
Schramm-Loewner class~\cite{smirnov01}. This leads to a characteristic
power law in the nodal domain sizes, which has been used to draw a
connection with critical percolation~\cite{bogomolny02,blum02}, in
which on any finite lattice a large spanning domain emerges. This
implies particular statistics for the density, areas and fractalities
of the domains~\cite{keating06}, which in turn are related to
conformal field theories that appear to characterise the spatial
structure of the eigenfunctions~\cite{keating06}. It is natural to
expect
that the conformations of random eigenfunctions in three dimensions
should be described by similar universalities, but it is not clear how
the statistics of a given field component should be modified by the
higher dimensionality as the analytical machinery to compare nodal
lines with models from statistical physics is less well developed.

\begin{figure*}
  \centering{
  \includegraphics[width=0.45\textwidth]{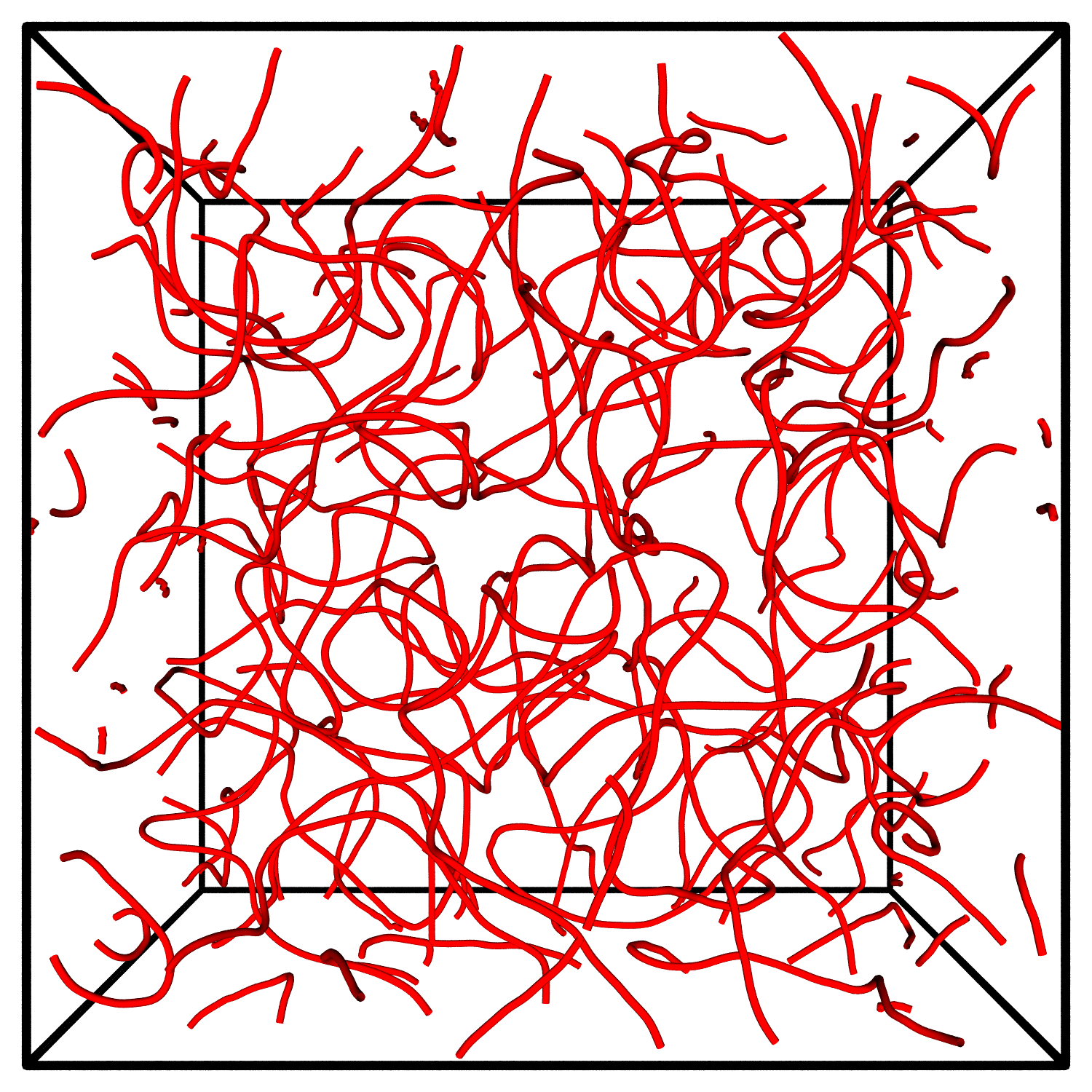}
  }
  \caption{Tangled vortices in an example random wave with periodic
    boundaries. This model is introduced in
    Section~\ref{sec:loopsoups}, shown here with energy $E=59$,
    periodic side length $3.84\lambda$. The tangle contains nine
    distinct vortex loops winding around one another.  }
    \label{fig:example_rwm}
\end{figure*}

In practice it is difficult to compute high energy eigenfunctions of
ergodic cavities even in two dimensions, and we draw our numerical
results for 3D wavefields, $\psi$, from the
\emph{random wave model}~\cite{berry77},
\begin{equation}
  \psi(\vec{r}) = \sum_n^N a_n \exp(i (\vec{k}_n \cdot \vec{r} + \chi_n))~,
  \label{eq:complex_rwm}
\end{equation}
for i.i.d Gaussian random complex $a_n$, uniform random phase
$\chi_n$, and randomly oriented $\vec{k}_n$. The sum over $N$ is in
principle infinite, but in practice the statistics of the wavefield
converge rapidly as $N$ grows larger than a few tens. When
$|\vec{k}_n|=k$ the random wave model is an excellent statistical
model for high energy eigenfunctions of the Laplacian,
$\nabla^2\psi = -E\psi$, in ergodic cavities far from the
boundaries~\cite{berry77}, and in two dimensions reproduces
statistical results including the random percolation-like behaviours
of nodal structures in quantum chaos as discussed above (in this case
the eigenfunctions can always be taken to be real, so the model is
equivalent to taking $\Re (\psi)$ or $\Im(\psi)$ as these are
independent).  However, \eqref{eq:complex_rwm} is not restricted to
two dimensions or even to the monochromatic spectrum, and can be a
statistical model for many different kinds of disordered wavefields
such as random interference in acoustic
resonances~\cite{newdirections} or optical volume
speckle~\cite{goodman76,oholleran08}, as well as 3D
resonances of chaotic cavities directly analogous to 2D
quantum chaos, although the complex waves here represent the breaking
of time-reversal symmetry such as in systems with absorption losses at
the boundaries.  Since~\eqref{eq:complex_rwm} is a continuous,
physical wave model, the nodal lines can be resolved at all scales,
from the smallest lengths at which their local spatial variation is
limited by the smoothness of the wavefield, to the larger lengthscales
beyond a certain cutoff where this behaviour `averages out' and they
behave as Brownian random walks. Regardless of the spectrum the
vortices then form a complex tangled loop soup, not morphologically
dissimilar to that in far more complex physical systems, despite the
randomness of the wavefield coming not from physical processes but
only the initial conditions of the
wavefunction~\cite{berry00,taylor14}. Figure \ref{fig:example_rwm}
shows a typical example of these tangled vortices in a 3-torus
eigenfunction with energy $E=59$; the model for this system is
described in Section~\ref{sec:loopsoups}.

In 3D complex wavefields, the vortex nature of the
nodal lines makes them especially privileged; the complex phase
circulates around them by $2\pi$ (in principle $2n\pi$ for any integer
$n$, but higher values are generically unstable), so they form a
structural skeleton to a space-filling structure of random phase
sheets. Much of the analysis of vortex filaments in 3D
complex random waves has focused on the counterparts of statistics
that in two dimensions are fixed by the power law scalings of critical
percolation. Certain statistical quantities, including the density of
vortex lines per unit volume and probability distribution of vortex
curvatures, have been computed analytically~\cite{berry00}. However,
other large scale quantities such as the fractality of the
vortices~\cite{oholleran08,taylor14} (both individually and in bulk),
and the distribution of their lengths~\cite{taylor14,taylor17}, have
been directly computed only numerically. These simulations generally
take place under periodic boundary conditions in order to be able to
follow individual vortex lines across large
lengthscales~\cite{oholleran08,taylor14}, which also allows for a
numerical analysis of tangling in terms of the vortex topologies, as
they may knot or link with one
another~\cite{oholleran09,taylor16}. The measurements have confirmed
that the vortices of these random tangles appear to have Brownian
random character resembling random walks at all lengths beyond some
small correlation lengthscale~\cite{taylor14,oholleran08}. On large
scales, the loop frequencies appear to follow a power law scaling
relation with respect to the loop lengths~\cite{taylor14,taylor17},
consistent with predictions for an ideal loop soup with no boundaries,
as previously observed in other systems such as cosmic
strings~\cite{taylor14}, and reminiscent of the power law for loop
length frequencies in 2D quantum chaos.

Other aspects of these results have appeared inconsistent with the
behaviour of an ideal loop soup. Extremely long vortex lines occur far
more frequently than the power law would indicate, usually wrapping
around the periodic boundaries a non-zero total number of times before
returning to their starting point. These wrapping lines have
\emph{non-trivial homology} (NTH), and would appear as infinite
(periodically repeating) lines in a tiling of space with the periodic
cell, but infinite lines are not compatible with the loop length
distribution in an ideal infinite loop soup~\cite{vachaspati84}. To
contrast with this behaviour, we sometimes refer to lines with trivial
homology as `closed loops', indicating that they do not ultimately
`wrap around' any periodic direction before closing.  The NTH lines
appear to be a manifestation of similar phenomena that are universal
in random lattice models~\cite{strobl97}, admitting an understanding
as a percolation-like phenomenon~\cite{hindmarsh95}, but the details of the
role in wave chaos were not fully quantified in previous
work~\cite{taylor14}. It is natural to expect that they may result
from the periodic boundaries as a finite-size effect, but their
frequency of occurrence has not appeared strongly related to the
periodic side length~\cite{taylor17}.

We show here that the scaling relations of vortex loop soups, across
all regimes from the wavelength scale to the large lengths where NTH
lines dominate, are predicted by recent work on the universal
statistics of vortex line tangle in short range correlated
fields~\cite{nahum12,nahum13,grosskinsky12}. In Section
\ref{sec:loopsoups} we show that the predictions of this model
accurately quantify the statistics of length distribution in random
wave vortices, as a direct 3D analogue to the
percolation-like statistics of nodes in two dimensions. These results
are demonstrated numerically via large scale simulations of vortex
tangle, in which the vortices are accurately tracked across all
lengthscales to accurately reconstruct the geometry of every
filament. In Section \ref{sec:reconnections} we use extensions to the
numerical model to follow vortex curves as they reconnect with one
another under different types of time evolution, and show that the
statistics of reconnection frequencies between vortices of different
lengths are strongly affected by the periodic boundaries. These
effects are well described by the predictions of~\cite{nahum13}, and
we show that the physical reconnection processes directly reproduce an
ideal statistical split-merge description of vortex line
dynamics~\cite{nahum12}.

\section{Loop soup statistics of vortex tangle}
\label{sec:loopsoups}

Following the literature~\cite{oholleran08,taylor14}, our model for
loop soups in wave chaos is not the ideal random wave model but the
cube with periodic boundary conditions, in which case the random wave
model is as in \eqref{eq:complex_rwm} but with $\vec{k}_n$ restricted
to wavevectors compatible with the side length of the periodic unit
cell. The valid solutions define \emph{arithmetic random waves} in
three dimensions~\cite{krishnapur13}, where for a given eigenvalue $E$
the valid wavevectors correspond to the intersections between a sphere
of radius $\sqrt{E}$ with integer valued lattice points. The number of
solutions is therefore finite, and not truly isotropic, but at
sufficiently high energies will contain enough degenerate components
to reproduce the statistics of the random wave
model~\cite{krishnapur13}. We draw our results primarily from the
monochromatic spectrum $|\vec{k}_n| = k$, with reference to the
wavelength $\lambda=2\pi/ k = 2\pi / \sqrt{E}$ as the only physical
lengthscale. Such a spectrum could reproduce the statistics of chaotic
eigenfunctions, but as the random statistics of loop soups depend
only on the isotropic random nature of the wavefield, all the
large scale results should generalise to any spectrum.
These arithmetic random waves can equivalently be considered as
degenerate random eigenfunctions of the Laplacian on the 3-torus, with
eigenvalue $E$, capturing the statistics of a deterministic
non-time-reversal-symmetric cavity. The advantage of using periodic
boundary conditions is that individual vortices can be followed along
their entire lengths rather than terminating on the edges of the
sampled cell, even if they wrap around them before closing and so form
NTH lines. In these cases their lengths are well defined even though
they present a topological problem when interpreted as loops.

The following results come from large scale simulations of the loop
soups in arithmetic random waves at each of $E=59$, $E=243$ and
$E=675$, whose degeneracies are $36$, $52$ and $112$ respectively. The
random waves at these energies also have a symmetry
$\psi(x + a\pi, y + b\pi, z + c\pi) = (-1)^{a+b+c}\psi(x, y, z)$,
implying that the nodal lines in each octant of the periodic cell are
identical, although the local signs of the real and imaginary field
components may be different. In the following analysis, we choose and
analyse a single unique octant of each eigenfunction, such that the
nodal lines have no additional symmetry (the same methodology was used
in~\cite{taylor14,taylor17}). The side length of the toroidal octant
is taken as $\pi$ or $L\lambda$ where $L=\sqrt{E}/2$, and is therefore
approximately $3.84\lambda$, $7.79\lambda$ and $13.0\lambda$ at each of
$E=59,243,675$.  The degeneracies of arithmetic random
waves at these energies are not especially high, but are the largest
available among similar energy scales, and appear to be high enough to
reproduce statistics of the random wave model~\cite{taylor14}. Some
nearby energies have a slightly higher degeneracy but lack the octant
symmetry; the unit cell of the vortex pattern becomes instead a
truncated octahedron, whose periodicities are more difficult to
implement numerically and so are not preferred here. The smallest
energy, $E=59$, is chosen because of the relatively small total vortex
length within such a small cell, allowing them to be trapped and
analysed relatively rapidly in the cases where statistics can only be
generated from many iterated simulations.

The vortices forming the loop soup are numerically located and tracked
using their local phase structure. The $2\pi$ circulation of the
complex phase can be detected by any phase integral around the vortex
core, even arbitrarily far from it, as long as no other vortices are
contained within the integral region. The full tangled vortex
structure is accurately reconstructed via the recursively resampling
Cartesian grid algorithm introduced in~\cite{taylor14,taylor17}, in
which vortex positions are first approximated at relatively low
resolution, then if necessary resampled to correct local tracing
errors (e.g. where vortices are detected to approach closely) and to
enhance local geometrical and topological precision. The base numerical
resolutions for the recovered vortex line representations are
$\sim 0.035\lambda$ at $E=59$ and $\sim 0.09\lambda$ at $E=243$ and
$E=675$, with the higher resolution at $E=59$ chosen to enhance the
resolution of reconnection events in the later analysis, although even
the lower resolution is well below the vortex correlation lengthscales
(geometrical decorrelation occurs on the scale of
$\sim \lambda$~\cite{taylor14}). In all cases, the local resolution is
increased by the algorithm where necessary to unambiguously capture
the vortex path, so these values are minimum sampling resolutions for
the recovered vortex curves. With these parameters, tracking the
vortices in a single random wave with $E=59$ takes $\sim 140s$ on a
modern laptop, while an $E=243$ eigenfunction takes
$\sim 80$s~\cite{taylor14} and an $E=675$ eigenfunction
$\sim 360$s~\cite{taylor17}.

The higher energy eigenfunctions $E=243$ and $E=675$, analysed by
these methods, have previously been established to closely reproduce
certain local statistics (densities, curvatures,
fractalities~\cite{taylor14}) of the isotropic random wave model, and
on large scales the the closed loops (but not NTH lines) have the
length distribution of a loop soup~\cite{taylor14,taylor17}. In order
to confirm that the numerical errors in recovering our vortex tangle
are comparable with these previous results, the same density
comparisons are as follows for eigenfunctions at $E=59$ (at $E=243$
and $E=675$ the numerical parameters exactly match those
of~\cite{taylor14}): the average density of nodal lines through an
axis-aligned plane is $(2.11\pm0.04)\lambda^{-2}$, compared to the
expected value of $2\pi/3\lambda^{2}\approx 2.09 \lambda^{-2}$ in the
isotropic model~\cite{berry00}, while the volume density is
$(4.59\pm0.02)\lambda^{-2}$ compared to the isotropic value
$4\pi/3\lambda^2 \approx 4.19 \lambda^{-2}$~\cite{berry00}. The planar
density matches the theoretical expectation well, while the volume
density is $\sim10\%$ higher. A comparable discrepancy was also seen
in~\cite{taylor14} even at higher energies, and was thought there to
arise from the torus periodicity encouraging vortex loops. Following
these previous results, we anticipate that these discrepancies should
not significantly affect most statistics, but an error on the order of
up to $\sim 10\%$ cannot be excluded from other numerically recovered
quantities. Since these results are no worse than at 
$E=243$ or $E=675$ (or even $E=1875$~\cite{taylor14}), we
anticipate that in fact the large scale results below are quite robust
to finite size effects beyond the immediate discrepancy in the
numerically recovered value.

A core scaling feature in ideal loop soups is that the filamentary
tangle should be fractally invariant across all lengthscales beyond
some small-scale cutoff, arising here from the wavelength scale, or
equivalently e.g. the lattice spacing in other models, analogous to
the percolation-fixed distribution of nodal domain areas in
2D quantum chaos~\cite{keating06}.  This requires a
particular power law in the scaling of loop lengths, with the number
of loops of length $l$ decaying as $l^{-\frac{3}{2}}$, the exponent
arising from the combination of Brownian fractality of the individual
vortices and the requirement that geometry of the tangle cannot depend
on the scale (a full derivation is given
in~\cite{vachaspati84,hindmarsh95}), or equivalently obtained as the
probability distribution for the lengths of random walks in three
dimensions that are conditioned to return to their
origin~\cite{scherrer86}.
This power law is therefore to be expected from the numerical
simulations described above, and has been found previously to hold for
the closed loops in the same arithmetic random wave
model~\cite{taylor14,oholleran08}, at lengthscales between
$\sim3.5\lambda$ (below which local correlations dominate) up to
almost the very largest loop scales. However, this counts only the
loops with trivial homology, and the NTH lines present in all of these
numerical experiments have not appeared consistent with this scaling
law.

\begin{figure}
  \centering
  \includegraphics{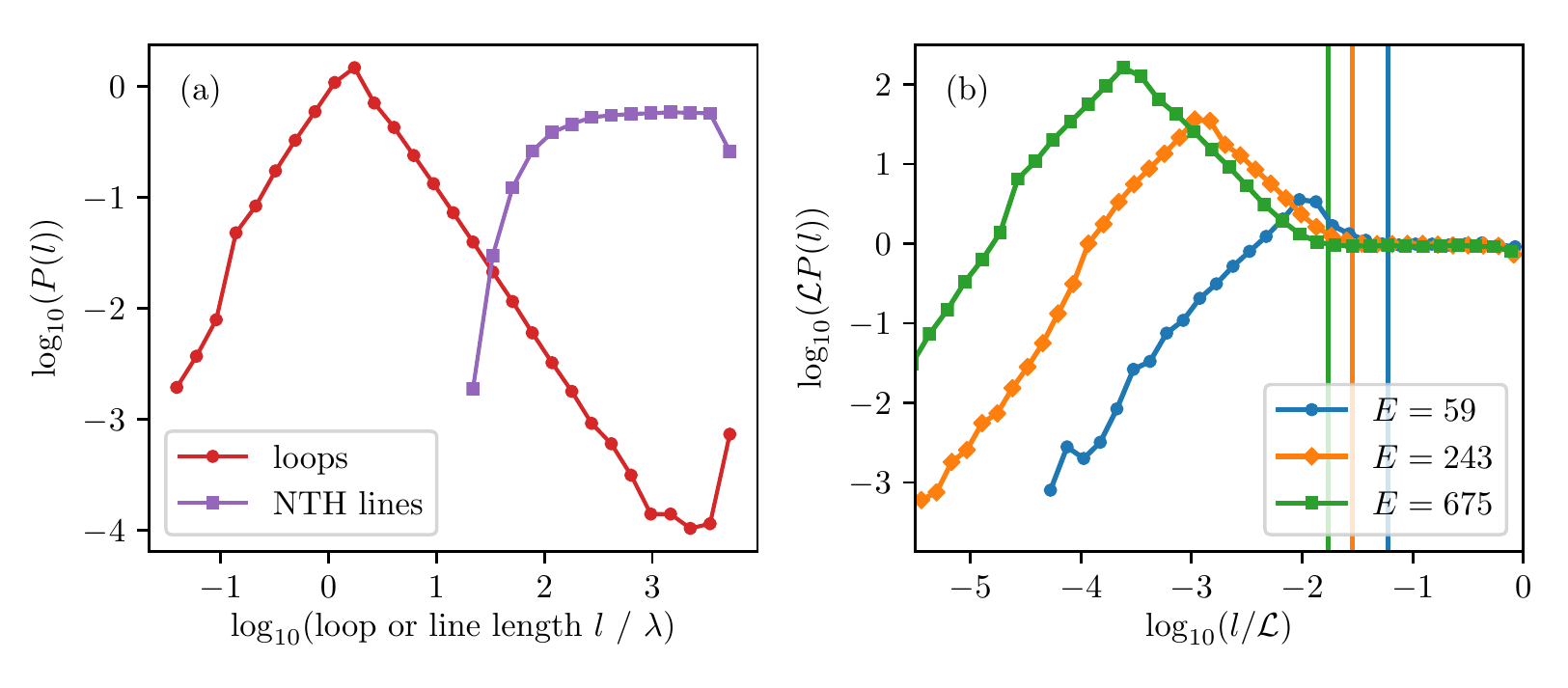}
  \caption{Probability distributions by length for loops and NTH lines
    in random waves with periodic boundaries. (a) shows on a log-log
    scale the probability distributions of vortex lengths for each of
    the $2583322$ closed loops (green) and $37745$ NTH lines (purple),
    normalised separately, taken from $9500$ different
    numerically-recovered random waves at $E=675$. (b) shows on a
    log-log scale the combined probability distributions of vortex
    lengths for both loops and NTH lines, for random waves with
    periodic boundaries at three different energies $E=59$ ($63114$
    vortices from $7157$ random wave simulations), $E=243$ ($1278186$
    vortices from $19845$ random wave simulations) and $E=675$
    ($2621067$ vortices from $9500$ random wave simulations, the same
    dataset used in (a)). Each vortex arclength is normalised by the
    total vortex arclength in its random wave. The peak in each case
    corresponds to $\sim3.5\lambda$. The vertical lines mark the
    lengthscale $L^2\lambda$ at each energy, where $L\lambda$ is the side length of
    the periodic cell, and correspond to the expected crossover from
    an ideal loop soup power law to a Poisson-Dirichlet distribution.}
 
    \label{fig:loop_and_line_scaling}
    \label{fig:pd}
\end{figure}

Figure~\ref{fig:loop_and_line_scaling}(a) shows this scaling regime as
expressed in arithmetic random waves with $E=675$. The loops exhibit a
power law scaling with gradient $-1.44 \pm 0.01$ across almost three
orders of magnitude, consistent with previous results in the same
model~\cite{taylor14}, and with the discrepancy from the anticipated
$-3/2$ thought to arise from the finite volume of the system.  The
distinctly different scaling of lines with non-trivial homology is
also shown, and these lines occur across all lengthscales larger than
the cell side length (the shortest possible length for a NTH
line). Despite their local geometrical similarity to closed loops, and
the fact that their wrapping number around the boundaries is generally
small compared to their total length, the NTH lines do not appear to
follow a power law in their length distribution. At lengthscales
beyond the cell side length they rapidly become much more numerous
than closed loops, but even in this regime the closed loops continue
to follow the loop soup power law across at least one further order of
magnitude, although this eventually breaks down and the loops become
unusually frequent again at length fraction $\sim1$. These longest
loops often consume over half the the arclength in the periodic
cell. This final peak appears to arise from the fact that there must
always be either 0 or at least 2 NTH lines in order for their wrapping
numbers vectors to add to 0, as by continuity the periodic cell cannot
accumulate phase circulation on its boundary. The peak in loop
probability appears to correspond to loops so long that the system
does not contain enough `spare' vortex arclength for an NTH line with
opposite wrapping to be likely to exist, so the relative likelihood of
extremely long loops compensates for missing NTH lines to preserve
their own, otherwise-flat distribution.  It is natural to expect that
the presence of NTH lines and their coupling to the loop distribution
may jointly be corrections to the scaling due to the periodic
`confinement' of the box, but previous analysis has found no clear
dependency on the energy of the eigenfunction~\cite{taylor17}.

The universal vortex statistics introduced in~\cite{nahum12,nahum13}
provide a unified perspective on the behaviour of both closed loop and
NTH line vortices~\cite{nahum12}
Following \cite{nahum13}, the distribution of vortex loop lengths
under periodic boundary conditions has a more complex form that
incorporates but is not fully described by the power law of an ideal
loop soup, with the vortices of length $l$ falling into three distinct
regimes. At low energies below a correlation length $\xi$, the loop
statistics depend on the system, here being limited ultimately by the
smoothness of the field on the scale of the wavelength. At scales
$\xi \ll l \ll L^2\lambda$ the loops are Brownian which implies a
$-d/2$ power law for loop length scaling in $d$ dimensions, consistent
at $d=3$ with the ideal loop soup model~\cite{vachaspati84}.
This regime represents those lines that are large enough not to be
dominated by local correlations but small enough not to wind through a
local volume larger than the periodic cell, such that they do not
`feel' its boundaries. The scaling with $L^2 \lambda$ arises from the
fact that the vortices, as random walks, have fractal dimension 2. In
general, these loops are not large enough to wrap around the
boundaries and approach periodic `copies' of themselves.  Although
this regime can be clearly observed in Figure
\ref{fig:loop_and_line_scaling} below the $L^2\lambda$ lengthscale,
the $-3/2$ power law continues in Figure
\ref{fig:loop_and_line_scaling}(a) for loops up to at least $50$ times
longer.

\begin{figure}
  \centering{
    \includegraphics[]{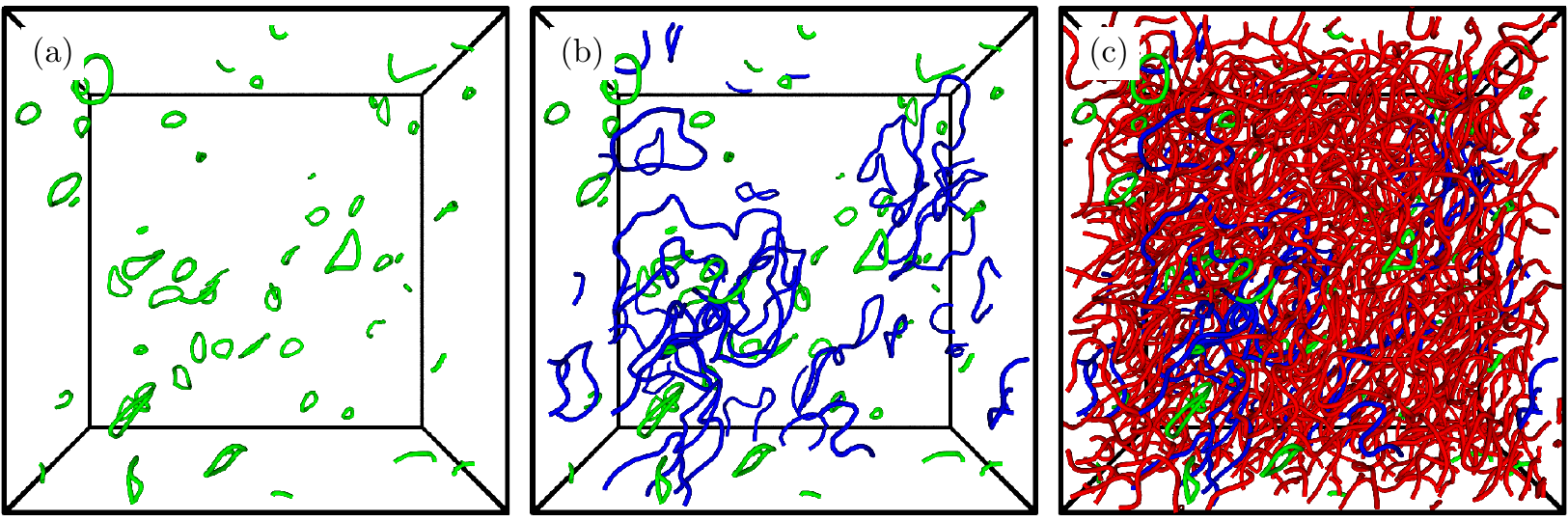}
  }
  \caption{Vortices across different lengthscales in a 3-torus
    eigenfunction with $E=243$ (side length $3.78\lambda$). (a) shows
    in green only the short vortices below the scaling cutoff
    $\xi\approx3.5\lambda$, which form perturbed
    ellipses~\cite{berry06}. (b) adds in blue the vortices in the
    regime $\xi < l < L^2\lambda$, i.e. large enough to take random
    conformations but smaller than the scale of the torus
    sidelength. (c) adds the vortices in the final length regime
    $l>L^2\lambda$, shown in red.  The short vortices in the green and blue
    regimes are most numerous, but the majority of total vortex
    arclength is contained within a small number of long vortices with
    $l\gg L^2\lambda$.}
    \label{fig:coloured_vortices_by_regime}
\end{figure}

Finally, at scales $l \gg L^2\lambda$, the combined distribution of loop and
NTH line lengths is expected to be Poisson-Dirichlet
(PD)~\cite{nahum13}. It should have the form
\begin{equation}
  P(l) = \frac{\theta}{\mathcal{L}} \left( \frac{1-l}{f\mathcal{L}}
  \right)^{\theta - 1}~,
  \label{eq:pd}
\end{equation}
where $\mathcal{L}$ is the total loop length in the system (here
$\mathcal{L}=4\pi L^3\lambda/3$~\cite{berry00}) and $\theta$ is a
\emph{fugacity} whose value depends on the number of loop types (which
may reconnect differently) and whether the loops are directed
(oriented). Vortices in wave chaos are oriented by their direction of
local phase rotation but have only one type, so $\theta=1$ and
\eqref{eq:pd} simplifies to $\mathcal{L}^{-1}$.

Figure \ref{fig:coloured_vortices_by_regime} shows how these regimes
are expressed in vortex loop soups via an example at $E=243$. (a)
includes only the short loops in the regime $l<\xi$, where the loops
are not large enough to behave as random walks and form instead
perturbed ellipses~\cite{berry06}. (b) adds the loops in the $-3/2$
fractal scaling regime, where the loops are large enough to behave
randomly, but not so large that they can fill the periodic cell. (c)
finally adds the longest lines, which are relatively few in number but
often consume most of the total arclength in the cell and usually have
non-trivial homology (in the Figure, every vortex in this regime has
NTH).

Figure~\ref{fig:loop_and_line_scaling}(b) shows the combined loop/NTH
linelength PDF for vortices at each of the investigated energies
$E=59,243,675$. In each case all three regimes can be clearly
distinguished. The crossover between small- and large-scale
behaviour occurs at $\xi \approx 3.5\lambda$, consistent at all
energies and with previous measurements~\cite{taylor14}. The position
of the later crossover to the Poisson-Dirichlet regime scales with 
$L^2\lambda$, as anticipated, marked here for each energy by a
vertical line, although this only approximately locates the crossover
as the transition between regimes is not perfectly sharp. The PDF is
flat beyond this point and $P(l)\propto \mathcal{L}^{-1}$, even at
relatively low degeneracies.

These results make it clear that the NTH lines arise from the periodic
`confinement', with their statistics depending directly on the
lengthscale at which a vortex line becomes long enough to `feel'
itself through the periodic boundaries. However, when $L$ grows
arbitrarily large, there will always be a small number of increasingly
long vortices in the Poisson-Dirichlet regime, even as the volume of
the periodic cell and number of closed loops in the fractal regime
grow rapidly. In fact the contributions of lines in the
Poisson-Dirichlet regime, which will usually have non-trivial
homology, rise at the same rate as the length contribution from closed
loops with $P(l) \propto l^{-3/2}$. According to these relative
frequencies, when $L$ is large the vortex arclength in closed loops
makes up on average $\sim 25\%$ of the total vortex length in a given
cell, although this result assumes a sharp transition between the two
regimes and that the closed loops follow a $-3/2$ power law even in
the Poisson-Dirichlet regime. It also does not count the vortex length
in the small loops with $l<3.5\lambda$, whose frequencies are not
described by the ideal model. However, as local phenomena their
frequency depends primarily on $L^3$, and in practice small loops appear to
contribute less than half as much vortex arclength as the closed loops
in the fractal regime. Accounting for all these approximations, we
anticipate that the closed loops in periodic random wave vortex tangle
with high $L$ will make up no more than $40\%$ of the total arclength
on average, with the rest of the arclength always expressed in a
relatively small number of lines with NTH. This loop length fraction
is much higher than the $~20\%$ seen in
practice~\cite{taylor14,oholleran08}, but these previous numerical
results all consider a relatively small $L$ in which the vortices do
not yet approach an ideal configuration. In particular, in these
simulations the ideal $-3/2$ power law scaling has not yet been
attained, significantly reducing the fraction of vortex length in this
regime.

\section{Vortex reconnections in dynamic systems}
\label{sec:reconnections}

The different statistical regimes discussed in Section
\ref{sec:loopsoups} are properties of the static vortex tangle. In
wave chaos the vortices have no physical dynamics and their random
statistics arise only from the random parameters of
\eqref{eq:complex_rwm}. However, the same result should arise from
time evolution in appropriate random systems. For instance, even in
linear isotropic wave superpositions that are not monochromatic
(i.e. the wavevectors have some spectrum in $k$), or fields with
time-varying phase (i.e. random waves
$\exp(i(\vec{k}\cdot\vec{r} + \omega t + \theta))$ for some frequency
$\omega$), the vortices are no longer static but will move
around~\cite{berry00}, sometimes \emph{reconnecting} with one another,
or shrinking to nothing as small loops, or likewise appearing from
nothing~\cite{berry06}. The distribution of vortex arclength must be
a statistical steady state under these processes.

The Poisson-Dirichlet regime can itself be understood statistically as the
result of a split-merge process between the different nodal lines, in
which individual vortex lines reconnect randomly such that the
limiting distribution of their lengths is
\eqref{eq:pd}~\cite{grosskinsky12,nahum13}. The steps of this
split-merge procedure are~\cite{grosskinsky12}:
\begin{itemize}
\item Choose a first vortex line $i_1$ with
  probability proportional to its length $l_1$.
\item Independently choose a second vortex line $i_2$ with probability
  proportional its length $l_2$ (this may be the same line).
\item If $i_1$ and $i_2$ are the same line, split the line to form two
  shorter lines with lengths $ul$, $(1-u)l$, for $u$ uniform random in
  [0, 1].
\item If $i_1$ and $i_2$ are different lines, join them to form a
  single longer line of length $l_1 + l_2$.
\end{itemize}
The Poisson-Dirichlet distribution is recovered in the limit of
repeated applications of this procedure. In physical systems the role
of the periodic boundaries is that they allow for long vortices to
roam throughout the box without being strongly locally biased. These
long lines in the Poisson-Dirichlet regime tend to approach every
point in the box, including passing close to all other regions of
themseves. Under random reconnections the resulting statistics are
therefore perturbed from the ideal loop soup power law despite the
local uniformity of the cell.

Since the reconnections in wave chaos between should be random events
depending only on the local geometry of the wavefunction, they may be
expected at large lengthscales to replicate the split-merge process of
the Poisson-Dirichlet distribution. It would then describe not just an
algorithm to reproduce the statistical distribution, but the actual
large scale behaviour of vortex tangle in wave chaos, even in systems
with e.g. very different spectra. Since these statistics are universal
to random vortex tangles, the same model should even describe the
local structure of configuration space for monochromatic waves;
perturbations to random wave parameters that preserve their random
statistical distributions will likewise cause vortices to move and
reconnect, with the same relative probabilities.

We investigate this reconnection behaviour numerically, via extensions
to the monochromatic random wave eigenfunctions previously
introduced. Under any change that causes the vortices to move, their
positions at successive time steps can be recovered by the RRCG
algorithm (see Section \ref{sec:loopsoups}), and reconnections
detected by their characteristic signature in the instantaneous change
in arclength distribution amongst the vortices; as long as the
vortices move smoothly, so does the set of vortex lengths, but
reconnections are seen as an instantaneous jump as either one
vortex splits into two components, or two vortices join
together. Detecting reconnections in this way is robust even when
multiple reconnections occur in a single numerical step, as long as
the changes in length of individual vortices can be distinguished from
the incremental changes of evolution without reconnection. In
practice, at accessible numerical resolutions almost all reconnections
can be distinguished from normal loop evolution in the length
spectrum, although reconnections involving the smallest loops may be
missed.  It would also be possible to further numerically isolate the
vortices involved in the reconnection, and to find the exact spatial
position of the event, but we do not do so below as it is not
necessary for the recovery of reconnection statistics.

\begin{figure*}
  \centering{
 
    \includegraphics{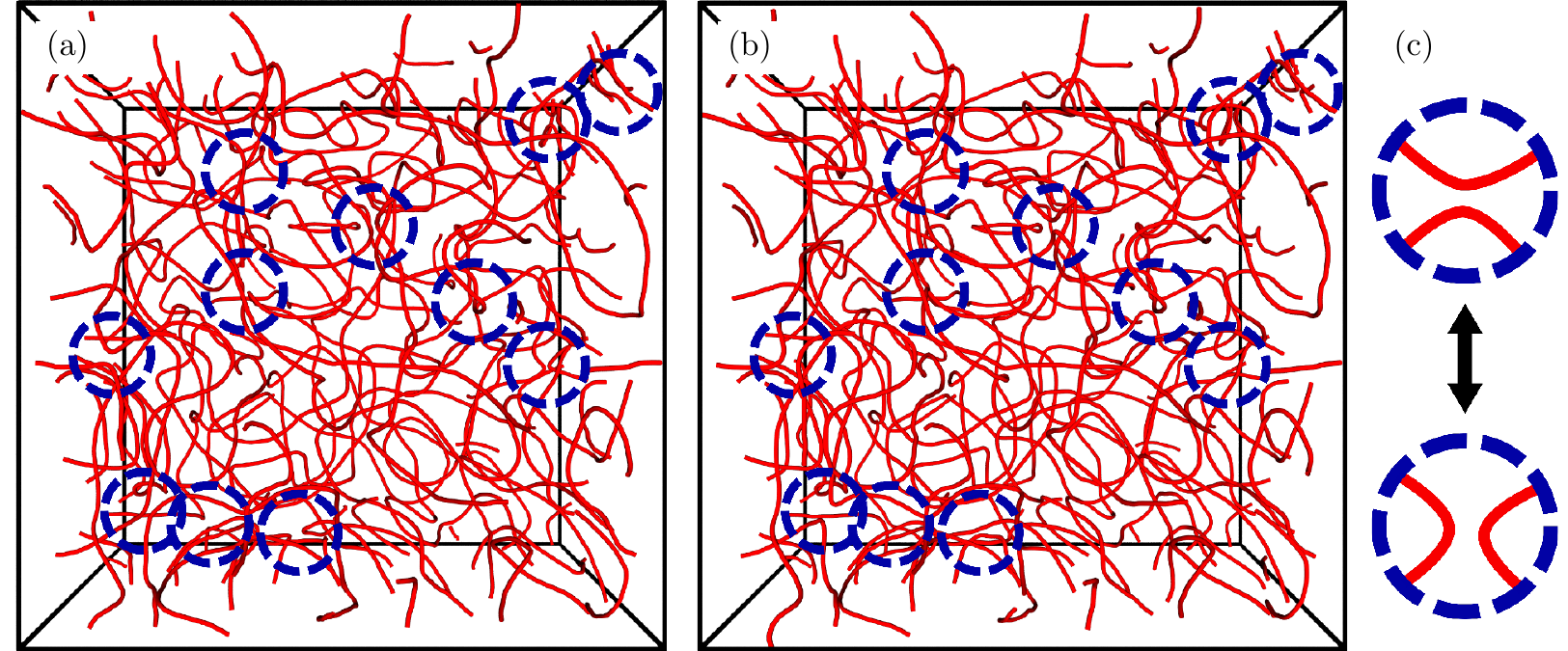}
  }
  \caption{Nodal lines in slightly different arithmetic random waves
    (3-torus eigenfunctions) with energy $E=59$. In (a), the
    amplitudes and phases or the 36 component plane waves are selected
    randomly. In (b) the eigenfunction is the same but with the phase
    of a single plane wave shifted by $\pi$. In most regions the local
    geometry of the nodal lines is barely affected by this change, but
    in several places (marked by blue circles) even this small
    adjustment induces a reconnection between vortex lines, changing the
    topology of the tangle. (c) illustrates how the vortices
    change locally during a reconnection; on the boundaries of the
    marked circle they remain essentially static, but in the centre
    they touch then split apart in a new configuration.

  }
    \label{fig:reconnection_comparison}
\end{figure*}

The expected reconnection rate in isotropic random wave systems has
recently been derived by~\cite{hannay17}, and this result can be used
to verify the numerical accuracy of our reconnection detection. An
accessible ensemble for comparison is the \emph{monochromatic modulus}
spectrum~\cite{hannay17}, in which the component plane waves of an
isotropic, monochromatic random wave model all simultaneously have
their phases changed at the same rate $\omega$, but each with random
sign. As with other random waves, this ensemble should be well
approximated by arithmetic random waves, as reconnections are
local events. We do so via 154 simulations of 3-torus eigenfunctions
with $E=59$, in each case assigning a random phase change sign to each
plane wave and incrementing in numerical phase steps
$\Delta \theta = 7.5\times10^{-4}$, chosen for numerical efficiency to
balance the probability of multiple reconnections occurring in the
same numerical step against the average number of steps between
reconnections. Each eigenfunction is incremented by up to $\sim785$
phase steps (total phase change $\sim0.59$), and its reconnections in
this range counted by detecting changes in the number of loops and by
observing discontinuities in instantaneous distribution of loop
lengths. This procedure does not distinguish between reconnections and
events where a single loop shrinks to nothing or appears from nothing,
although the birth and death of loops appears relatively uncommon,
consistent with~\cite{hannay17} in which its rate is $\sim26.5$ times
lower than that of reconnection. Counting $17156$ separate
reconnection events over a total phase change $\Delta\theta=77.49$,
the average combined rate of reconnections rate is
$3.82 \lambda^{-3} s^{-1}$, compared to an expected rate of
$3.91 \lambda^{-3} s^{-1}$~\cite{hannay17}. The discrepancy in these
numbers is consistent with those in other quantities under periodic
boundary conditions, especially as it is likely that our numerical
method slightly under-counts the reconnection rate due to being
unable to detect reconnections involving very small loops, where the
length change is below a normal length change during evolution.
It is also possible that the reconnection rate is suppressed by the same
finite-size effects that slightly reduce the volume density of vortex
lines. We therefore consider that this technique for detecting
reconnections is numerically reliable.

We investigate many reconnection events by numerical experiments using
3-torus eigenfunctions with $E=59$, chosen because the vortices in
individual eigenfunctions are few enough to be rapidly tracked,
allowing the change in the vortex tangle to be followed efficiently
over the course of significant perturbations to the random state.  A
given random eigenfunction with degeneracy $\mathcal{D}$ is the sum of
$\mathcal{D}$ plane waves, and therefore $\mathcal{D}$ Gaussian random
amplitudes and $\mathcal{D}$ uniform random phases; here,
$\mathcal{D}=36$. If each of these parameters undergoes a random walk,
appropriately rescaled such such that coefficients retain their
original random distributions, vortices will move and reconnect
randomly, but representing the random structure of the local
configuration space rather than explicit time dynamics.  As an even
simpler model that should reproduce the same reconnection statistics,
we take at random a single plane wave component of the random
eigenfunction, and slowly increment its phase until a reconnection
occurs; a phase step of $\pi/100$ gives sufficient resolution. Even
with such a small perturbations, reconnections usually occur within 5
or fewer phase steps.
In each case, the vortices taking part in the reconnection are
recorded, the eigenfunction discarded, and a new set of random
parameters selected.  The choice to modify a single phase is simpler
than updating all random coefficients simultaneously, and does not
require an additional renormalisation to preserve the random
distribution of amplitudes. Since reconnections occur even at small
perturbations, and the configuration is not reused, we do not
anticipate that the methodology should bias reconnection statistics.

\begin{figure}
  \centering{
    \includegraphics{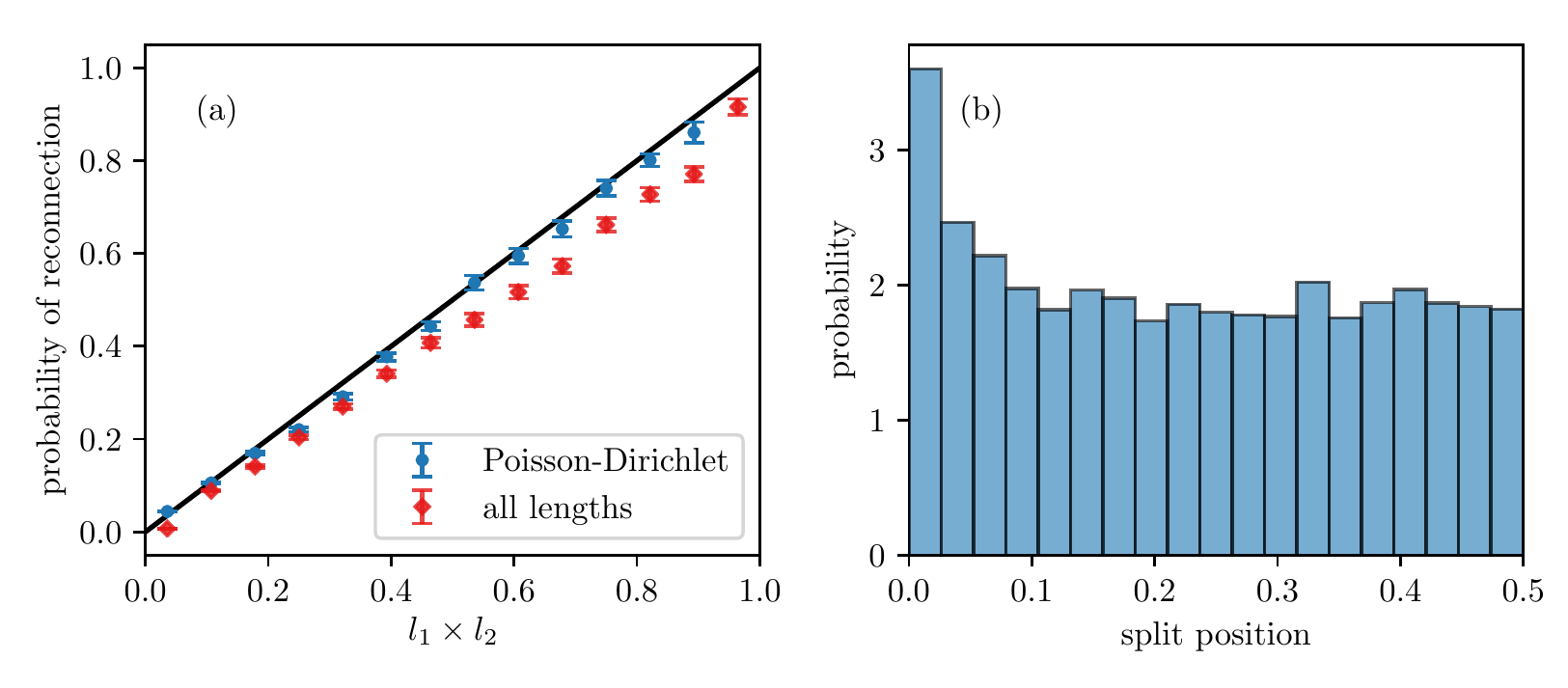}
  }
  \caption{Statistics of arclength redistribution during splitting and
    merging of random wave vortices. (a) gives probabilities of
    reconnections between nodal lines of different lengths. In the
    x-axis, reconnection events are binned by the multiple $l_1 l_2$
    of their length fractions $l_1$ and $l_2$; in the
    Poisson-Dirichlet model, this represents the probability of these
    lines being involved in the next reconnection event. The vertical axis
    counts the probability of these lines reconnecting (or a single
    line splitting) in the next reconnection event, found by taking a
    random wave under periodic boundary conditions and smoothly
    increasing the phase of a single plane wave component, stopping at
    the moment of the first reconnection as detected via the
    instantaneous jump in line length distribution.  Assuming the
    reconnections obey the split-merge statistics of the
    Poisson-Dirichlet distribution, both axes should represent
    the same quantity and the distribution will be fit by a straight
    line with gradient 1. The blue points show these results for 15865
    reconnections amongst lines in the Poisson-Dirichlet length regime
    ($l_1,l_2 \gg L^2\lambda$), where the statistics fit the expected
    straight line. The red points show the same results for 21574
    reconnections including vortices of all lengths, but as expected
    do not fit to the split-merge prediction. (b) shows the statistics
    of the positions of reconnections that split vortex lines in the
    Poisson-Dirichlet regime. The horizontal axis marks the length
    fraction along the original line at which the next reconnection
    takes place, while the vertical axis shows the (normalised)
    relative frequency of these reconnections. The results come from
    9515 independent vortex reconnection events at $E=59$, including
    only vortices whose initial lengths exceed 0.06 of the total
    vortex arclength in their eigenfunction.  }
   
  \label{fig:reconnection_probabilities}
  \label{fig:split_probabilities}
\end{figure}

Figure \ref{fig:reconnection_comparison} illustrates the instability
of local vortex topology under this single phase perturbation,
comparing two instances of similar eigenfunctions at $E=59$; the
function of (b) is the same as (a) but with one of the 36 random
phases shifted by $\pi$. Although the bulk of the vortex geometry is
hardly changed, this small perturbation results in the 11 marked
reconnection events, as well as others that cannot be seen as they
have already occurred and reversed within the $0\rightarrow\pi$
shift. In fact this example represents changing the phase of a
relatively low-amplitude wave; in other cases it is possible to induce
many more reconnections under the same $\pi$ phase shift.

The resulting statistics of vortex lengths during reconnection are
summarised in Figure \ref{fig:reconnection_probabilities}(a), for
$22150$ unrelated reconnection events from eigenfunctions with
different random configurations. The dark points mark reconnection
probabilities for lines in the PD regime only (here, length fraction
$\geq 0.06$ following the cutoff of Figure \ref{fig:pd}, although the
result is robust to different choices). This includes $15259$ of the
$22150$ detected reconnection events. The plot compares the multiple
of the vortices' normalised lengths, $l_1 l_2$, with the
numerically-recovered probability of their reconnection as compared
against all the other possible reconnections for lines in the PD
regime. If the split-merge procedure precisely describes the
statistics of vortex reconnection, these quantities should be the
same, and the results fit to a straight line with gradient $0.5$.
This result appears to match the behaviour of the vortex lines, and
the split-merge process is a good descriptor of vortex reconnection
rates in this regime. In contrast, the red points count all numerical
reconnections, including those involving lines below the PD length
regime. It is clear that the PD reconnection statistics no longer
apply, consistent with these lines being too short to interact
uniformly with all others.

The split-merge process also requires that when a PD-regime line
reconnects to split into two shorter lines, the vortex arclength
should be distributed with uniform probability amongst the two new
curves~\cite{grosskinsky12} (i.e. the reconnection is randomly
positioned). The actual distribution of these length split fractions,
recovered numerically from the simulations, is shown in Figure
\ref{fig:split_probabilities}(b) for the 9515 independent vortex
reconnection events in which a vortex line in the Poisson-Dirichlet
regime undergoes a split. As expected, the split position appears
uniform within statistical errors from the limited dataset, but only
beyond the Poisson-Dirichlet length cutoff of
$\sim0.06\mathcal{L}$. Reconnections below this scale are more common,
and represent small loops breaking off the Poisson-Dirichlet line,
often with lengths even below even the fractal behaviour cutoff of
$\sim3.5\lambda$ (see Section \ref{sec:loopsoups}). This may result
from the smoothness of the wavefield limiting the random behaviour of
the vortex lines on small scales, so that the formation of small
loop-like structures is encouraged and the reconnections in this local
region are relatively likely to germinate a new small loop. This
deviation from the Poisson-Dirichlet split-merge statistics is to be
expected when smaller loops are involved, and does not affect the
overall distribution.

\section{Discussion}

We have demonstrated that the soup of filamentary vortex loops in wave
chaos, sampled via large scale numerical simulations with periodic
boundaries, falls into the universal scaling classes recently
described by~\cite{nahum13}. These results are 3D
analogues to the well-studied percolation-like statistics of nodal
domains in 2D quantum chaos. Importantly, they capture
the effect of dimensionality on the vortex behaviour, as the wave
vortices can now wind around one another in three dimensions. This
result explains previous numerical observations of random wave
vortices~\cite{oholleran08,taylor14}, in which a deviation of the
vortex tangle from ideal loop soup statistics was observed but not
related directly to the impact of the periodic
boundaries~\cite{oholleran08,taylor14}. The long vortex lines with
non-trivial homology, which also arise generically in other classes of
random filaments~\cite{hindmarsh95}, are seen to arise naturally from
the periodic boundaries. Although they can be considered as a type of
boundary effect, they would always be present regardless of the
periodic side length, with their contribution to the total vortex
arclength balanced by the increased number of smaller, closed loop
vortices.
The length distribution of these long lines can be understood in terms
of a statistical split-merge process, and by observing the changes in
vortex lengths under perturbations to the configuration (or
equivalently, under time evolution in non-monochromatic spectra) we
have shown that this statistical process directly describes the
physical reality of how vortices reconnect with one another under
evolution processes. Although our analysis has focused on the
monochromatic spectrum, such as would describe the chaotic
eigenfunctions of 3D ergodic cavities, the same scaling
results should apply to any isotropic spectrum even where this changes
the local statistics dramatically.

The different statistical regimes of fractal loop soup behaviour can
be understood in terms of a correction to the random reconnection
statistics of an infinite random tangle, as a result of the periodic
boundaries.  This description is summarised as follows: the most
important property of the vortex lines is that beyond some small
correlation lengthscale they behave as closed Brownian random walks
(although they are self-avoiding according to local correlation
functions, the isotropic soup of other lines in all directions means
that their large scale directionality is not
biased~\cite{hindmarsh95}). In the absence of periodic boundaries,
this fixes their loop soup behaviour for all lengthscales above the
small scale cutoff, so that all vortices form closed loops, and
arbitrarily large loops can occur but only with low relative
probabilities.  Under perturbations to the state, or time evolution in
polychromatic wavefields, the vortices will move and undergo
reconnection events, but these are uniformly distributed about the
tangle and the reconnections preserve the ideal loop soup
distribution. The reconnections depend only on certain local
statistics, and are not directly affected by periodicity, but the
topology of the periodic cell changes the global statistics of how
vortices approach one another, as long vortices can now wind
throughout the periodic cell without necessarily being locally biased
any more than in isotropic flat space. These highly extended vortices
are long enough that they regularly pass close to other segments of
themselves, as well as to almost all the other loops in a given cell,
and the statistics of their reconnections are proportionately
affected. The result is the ideal Poisson-Dirichlet distribution, in
which the statistics of reconnections are simplified by the long loops
filling the cell so uniformly that only their relative lengths
contribute to the probability of their reconnecting.

It is interesting that despite the clearly delineated scaling regimes,
the ideal loop soup power law scaling persists for loops with trivial
homology even at lengthscales well into the Poisson-Dirichlet regime,
with the Poisson-Dirichlet statistics made up instead by the NTH
lines. Despite this clear separation of behaviours, the wrapping
numbers of the NTH lines around the periodic boundaries (i.e. homology
vectors) tend to be extremely small compared to the arclength of the
entire line, such that the difference between NTH lines and closed
loops would otherwise appear geometrically unimportant.

The reconnections are here considered only in a statistical sense, but
previous work in the literature has established the local behaviour of
the wavefield during such events; the vortices form hyperbolas, and at
the moment of reconnection a critical point of the intensity (i.e.
$\nabla |\psi|^2 = 0$) passes through the intersection point of the
vortex strands~\cite{berry06,dennis08}. Analytically describing the
statistics of these points remains an open problem~\cite{taylor14},
but it is tantalising that these numerical results, and recent work
calculating reconnection rates~\cite{hannay17}, give specific bounds
on how frequently their interaction with vortices becomes
important. It is possible that our numerical techniques could be
enhanced to capture the behaviour of these critical points and further
refine the understanding of reconnection events. These phenomena also
relate to other types of filamentary defects in complex
3D random waves; the intensity critical points sit on
lines along which the vorticity
$\frac{1}{2} \mathrm{Im}(\nabla \psi^* \times \nabla \psi)$ vanishes.
These are more difficult to detect than vortex lines, as they lack a
simple phase parameter whose accumulation can be detected, but they
must wind through the field with their own random
statistics~\cite{dennis08}. These lines are unoriented, which is
expected to change the parameter $\theta$ in \eqref{eq:pd} by affecting
the way the lines can reconnect, and leading to different loop soup
statistics within any Poisson-Dirichlet regime. Lines of this type
also occur as e.g. the L lines in 3D polarisation
fields~\cite{dennis08}, so it is important to note that the specific
large scale behaviour of (oriented) vortices is not completely generic
amongst other random filaments in similar fields.

Although reconnections instantaneously redistribute vortex arclength
amongst the different vortices of the loop soup, they do not
ultimately affect local geometrical statistics such as the average
curvature and torsion of the curves. A different kind of measurement,
sensitive to the large scale effects of these changes, is to ask
instead about the topology of the vortex loops; whether they are
knotted, or link with one another.  It has been established that
random waves at high energies generically contain
knotted~\cite{taylor16,taylor17} and
linked~\cite{oholleran09,taylor17} vortices. These topological
quantities measure the large scale entanglement of the loop soup, as
they depend on the global winding of the loops about themselves on
large scales, which may be quite decoupled from their local
geometrical correlations. This topological sensitivity to large scale
conditions has been seen in previous work where the statistics of
vortex knotting vary significantly amongst different systems of random
waves even at high energies where the local geometry is similar, due
to the knot types feeling the effect of their boundary
conditions~\cite{taylor16}. Since knotting is common at high energies,
the wave systems must exhibit some persistence of topological
complexity under reconnection, in contrast to the behaviour of the
small vortex knots of specific types that have been constructed in
other systems, but which usually break down quickly under
reconnections to become topologically trivial~\cite{kleckner13}. It is
possible that the right choice of topological quantity would reach a
statistical steady state under reconnection processes, like the
fractal and Poisson-Dirichlet regimes discussed above, but being
instead a type of topological universality.

Although topological quantities are a natural choice for measuring the
large scale entanglement of the system, it is not clear what
measurements would be appropriate to quantify topological complexity.
Properties of the knots and links can be captured using topological
invariants, functions of the curve whose value depends only on its
knot type, but these invariants are often abstract quantities without
a simple physical interpretation~\cite{taylor16}, and their values may
not be directly useful as descriptors of the vortex tangle as they can
change significantly during a single reconnection between
vortices. Attempts to quantify the topology of periodic systems would
also be complicated by the NTH lines which, as they take advantage of
the periodic boundaries, cannot be described as normal knots in
$\mathbb{R}^3$ (where they are naturally represented as infinite
periodically-repeating lines). These problems have been resolved in
the literature for the certain measurements of linking between curves
in periodic boundaries~\cite{panagiotou15}, but this method alone does
not detect the knotting of a single curve, and still does not rule out
large changes in the linking calculation as a result of even a single
reconnection event. It is therefore an open question whether the
vortex tangle will display topological universalities analogous to the
scaling results discussed above, or whether different types of loop
soups will exhibit characteristic differences in their knotting and
linking despite their consistent large scale fractality.

\ack

We are grateful to S G Whittington, R Evans, J H Hannay, and especially to M R
Dennis, for useful discussions. This research was funded by the
Leverhulme Trust Research Programme Grant No. RP2013-K-009, SPOCK:
Scientific Properties of Complex Knots.  This work was carried out
using the computational facilities of the Advanced Computing Research
Centre, University of Bristol.

\bibliography{papers}

\end{document}